\documentclass[aps,prl,twocolumn,nofootinbib,tightenlines,preprintnumbers,amsmath,amssymb,notitlepage,10pt]{revtex4-1}

\usepackage{epsfig}
\usepackage{latexsym}
\usepackage{booktabs}
\usepackage{color}
\usepackage[dvipsnames]{xcolor}
\usepackage{footnote}
\usepackage{hyperref}
\usepackage{simplewick}
\usepackage[caption=false]{subfig}
\usepackage{bm}
\usepackage{York}
\newcommand{\ETMCold}{Wagner:2011ev,*Bicudo:2012qt,*Bicudo:2015vta}
\newcommand{\oldlatt}{Richards:1990xf,*Mihaly:1996ue,*Green:1998nt,*Stewart:1998hk,*Michael:1999nq,*Pennanen:1999xi,*Cook:2002am,*Detmold:2007wk,*Bali:2011gq,*Brown:2012tm}

\begin{document}

\title{Lattice Prediction for Deeply Bound Doubly Heavy Tetraquarks}
\author{A. Francis$^a$, R. J. Hudspith$^a$, R. Lewis$^a$ and K. Maltman$^{b,c}$} 

\affiliation{
$^a$Department of Physics \& Astronomy, York University, 
Toronto, ON M3J 1P3, Canada\\
$^b$Department of Mathematics \& Statistics, York University, 
Toronto, ON M3J 1P3, Canada\\
$^c$CSSM, University of Adelaide, Adelaide SA 5005, Australia
\vspace{0.2cm}}

\date{\today}

\begin{abstract}
We investigate the possibility of $qq^\prime \bar b \bar b$ tetraquark bound states using $n_f=2+1$ lattice QCD ensembles with pion masses $\simeq 164$, $299$, and $415$ MeV. Motivated by observations from heavy baryon phenomenology, we consider two lattice interpolating operators both of which are expected to couple efficiently to tetraquark states: one with diquark-antidiquark and one with a meson-meson structure. Using nonrelativistic QCD to simulate the bottom quarks, we study the $ud\bar b \bar b$, $\ell s\bar b \bar b$ channels with $\ell=u,d$, and find unambiguous signals for strong-interaction-stable $J^P=1^+$ tetraquarks. These states are found to lie $189(10)$ and $98(7)$ MeV below the corresponding free two-meson thresholds.
\end{abstract}

\maketitle

\section{Introduction}
In QCD the attractive nature of the color Coulomb
potential for two antiquarks in a $3_c$ color configuration ensures
the existence of strong-interaction stable $qq^\prime \bar{Q}\bar{Q}$
exotics in the limit $m_Q\rightarrow\infty$~\cite{Carlson:1987hh,Manohar:1992nd}.
However, the physical charm and bottom masses are insufficiently
heavy for this mechanism to dominate and ensure the existence of
doubly charmed or bottom tetraquarks. Whether such exotics
exist is thus a dynamical question involving other, in general
non-perturbative, effects in these systems.

The spectrum of bottom baryons suggests that doubly 
bottom tetraquarks should exist and be strong-interaction stable,
for the following reasons.

First, the smallness of the observed $B^*-B$, $B_s^*-B_s$, $\Sigma_b^*-\Sigma_b$ 
and $\Xi_b^*-\Xi_b^\prime$ splittings suggests that $m_b$ is sufficiently
large that the heavy quark limit, in which the heavy quark spin decouples
and a heavy anti-diquark behaves like a single heavy quark, is a
reasonable approximation for $b$ quarks. 
The $\Sigma_b-\Lambda_b$ and
$\Xi_b^\prime -\Xi_b$ splittings then give a direct experimental measure
of the difference in energies between dressed diquarks with light ($u,d,s$)
quark flavor, spin and color $(F,S_\ell ,C)=(\bar{3}_F,\, 0,\, \bar{3}_c)$
and $(6_F,\, 1,\, \bar{3}_c)$ in the field of a heavy, nearly static
color $3_c$ source. The $\bar{3}_c$ non-strange $(I=0,\, S_\ell = 0)$
and $(I=1,\, S_\ell = 1)$ diquarks lie $\sim 145$ MeV
below, and $\sim 48$ MeV above, the corresponding spin average, while
the $\bar{3}_c$ $us$, $ds$ $(F,\, S_\ell )=(\bar{3}_F,\, 0)$
and $(6_F,\, 1)$ diquarks lie $\sim 106$ MeV below and $\sim 35$ MeV
above the corresponding spin average. Following Jaffe~\cite{Jaffe:2004ph}
we refer to the $(\bar{3}_F,\, 0,\, \bar{3}_c)$ and 
$(6_F,\, 1,\, \bar{3}_c)$ configurations as 
``good diquark'' and ``bad diquark''.

In the doubly heavy $qq^\prime \bar{b}\bar{b}$ sector, with $q,q^\prime =u,d,s$
the good $qq^\prime$ diquark configuration is accessible when the
two $\bar{b}$'s are in a color $3_c$. With  
no $\bar{b}$-$\bar{b}$ spatial excitation, the
$\bar{b}\bar{b}$ spin will be $J_h=1$, producing an associated 
tetraquark configuration with $J^P=1^+$. The relevant two-meson thresholds,
below which such a tetraquark would be strong-interaction stable, are
then $BB^*$ and $B_sB^*$ for the $I=0$ and $I=1/2$ members of the
$\bar{3}_F$, respectively. The residual spin-dependent interactions
in the $J=0,1$ mesons making up these threshold states are
suppressed by the $b$ mass, the physical thresholds lying
$23$ and $25$ MeV below their corresponding spin-averaged
versions. With $\sim 20$ MeV of further binding likely from the 
color Coulomb interaction in the $3_c$ $\bar{b}\bar{b}$ pair, one
would expect an $I=0$, $J^P=1^+$ $ud\bar{b}\bar{b}$ tetraquark bound by
considerably more than $100$ MeV and a related $J^P=1^+$
$\ell s\bar{b}\bar{b}$, $\ell =u,d$, isodoublet bound by $\sim 100$ MeV, in the limit
that one ignores the non-pointlike nature of the nearly static
$\bar{b}\bar{b}$ color $3_c$ source.

Expectations are less clear for the $qq^\prime\bar{c}\bar{c}$ tetraquark
channels, where spin-dependent light-heavy interactions are 
less suppressed. Phenomenologically, the $\Sigma_c^*-\Sigma_c$,
$\Xi_c^*-\Xi_c^\prime$, $D^*-D$ and $D_s^*-D_s$ splittings are a
factor three greater than the corresponding bottom sector splittings,
and the $\Sigma_c-\Lambda_c$ and $\Xi_c^\prime -\Xi_c$ splittings, which
would reflect purely light-quark spin-dependence in the heavy-quark limit, 
are $\sim 30$ MeV less than the corresponding bottom sector splittings. 
The larger ($\sim 140$ MeV) $D^*-D$ and $D^*_s-D_s$ splittings also
drive the $DD^*$ and $D_sD^*$ strong-decay thresholds farther down
below their spin-averaged analogues than is the case in
the bottom system, further eating into the potential tetraquark binding
generated by the good light diquark configuration. With significantly
reduced net spin-dependent attraction, other residual effects
will become more important to take into account quantitatively in the
double-charm sector.

Many studies in the literature argue for the existence of strong interaction stable doubly heavy tetraquarks \cite{%
Manohar:1992nd,Carlson:1987hh, 
Jaffe:2004ph, 
Ader:1981db, 
Tornqvist:1993ng, 
Richards:1990xf,*Mihaly:1996ue,*Green:1998nt,*Stewart:1998hk,*Michael:1999nq,*Pennanen:1999xi,*Cook:2002am,*Detmold:2007wk,*Bali:2011gq,*Brown:2012tm, 
Wagner:2011ev,*Bicudo:2012qt,*Bicudo:2015vta,
Ikeda:2013vwa, 
Guerrieri:2014nxa,Bicudo:2015kna, 
Zouzou:1986qh,*Lipkin:1986dw,*SilvestreBrac:1993ss,
*Semay:1994ht,*Pepin:1996id,*Brink:1998as,*Barnes:1999hs,*Gelman:2002wf,
*Vijande:2003ki,*Janc:2004qn,*Ebert:2007rn,*Vijande:2007rf,*Zhang:2007mu,*Lee:2009rt,*Vijande:2009kj,*Yang:2009zzp,*Valcarce:2010zs,*Carames:2011zz,*Ohkoda:2012hv,
*Hyodo:2012pm,*Silbar:2013dda}, the majority of these, moreover, identifying the $\bar{3}_F$, $J^P=1^+$ channel as optimal for binding.
The long distance $J^P=1^+$ $DD^*$ and $BB^*$ interactions,
mediated by pseudoscalar exchange, and hence constrained by chiral symmetry,
are known to be attractive~\cite{Manohar:1992nd,Tornqvist:1993ng}, 
with a strength capable of producing weak binding in the 
$BB^*$ system~\cite{Manohar:1992nd}.
Doubly heavy tetraquark channels have also been investigated in a number
of lattice studies~\cite{\oldlatt,Wagner:2011ev,*Bicudo:2012qt,*Bicudo:2015vta,Ikeda:2013vwa,Guerrieri:2014nxa,Bicudo:2015kna}.
The simulations 
all have $m_\pi >330$ MeV, and, 
with the exception of Ref.~\cite{Guerrieri:2014nxa}, treat the heavy
quarks in the static limit. All find the strongest attraction in the
$I=0$, $J^P=1^+$ channel. Evidence of increasing attraction with
decreasing light quark mass is also found. 
Refs.~\cite{\ETMCold,Ikeda:2013vwa,Bicudo:2015kna}
employ a Born-Oppenheimer approach, fitting static energies as a function of
heavy diquark separation, $r$, to an assumed functional form, $V(r)$,
and then solving for the heavy quark motion using the Schr\"odinger equation
with potential $V(r)$. The most recent of the studies employing
$n_f=2$ ETMC ensembles, Ref.~\cite{Bicudo:2015kna}, uses a screened 
color Coulomb form
for $V(r)$, and, extrapolating linearly to physical $m_\pi$, obtains an
estimated binding of $90^{+43}_{-36}$ MeV in the $ud\bar{b}\bar{b}$ channel.
Bound doubly bottom states, and sometimes bound doubly charmed states,
were also obtained in calculations employing a range of model-dependent
effective quark-quark interactions~\cite{Carlson:1987hh,Zouzou:1986qh,*Lipkin:1986dw,*SilvestreBrac:1993ss,
*Semay:1994ht,*Pepin:1996id,*Brink:1998as,*Barnes:1999hs,*Gelman:2002wf,
*Vijande:2003ki,*Janc:2004qn,*Ebert:2007rn,*Vijande:2007rf,*Zhang:2007mu,*Lee:2009rt,
*Vijande:2009kj,*Yang:2009zzp,*Valcarce:2010zs,*Carames:2011zz,*Ohkoda:2012hv,
*Hyodo:2012pm,*Silbar:2013dda}. 
The effective interactions employed in these studies all produce a
good diquark-bad diquark splitting compatible in sign and
magnitude with that required for understanding the features of
the meson and baryon spectra.

In what follows, we investigate more quantitatively
the expectation that a $\bar{3}_F$ of doubly heavy strong-interaction-stable tetraquark states exists via lattice calculations. 
Since the arguments above suggest that
binding will be greatest in the bottom sector,
and will increase as the light quark mass is decreased, we focus on
the $ud\bar{b}\bar{b}$ and $\ell s\bar{b}\bar{b}$ channels, and employ
publicly available $n_f=2+1$ ensembles with sufficiently low $m_\pi$ to allow for a controlled chiral extrapolation.
The ensembles used have near-physical $m_K$ and $m_\pi \simeq 164$ MeV, $299$ MeV and $415$ MeV. 
We work at fixed lattice
spacing, and use NRQCD for the $b$ quarks. 
With significant binding expected, a simple, well-adapted
$\bar{3}_F$, $J^P=1^+$ operator choice should suffice to extract
the ground state tetraquark signal. 
In fact, we employ two operators, one having the
diquark-antidiquark structure suggested by the discussions above,
and another whose flavor-spin-color correlations are those of the
two-meson $BB^*$ and $B_sB^*$ threshold states. Including the latter allows 
us to reduce excited state contamination of the ground state signal at
earlier Euclidean times through a $2\times 2$ generalized eigenvalue
problem (GEVP) analysis.
The results of our analysis bear out the expectations outlined above,
producing strong evidence for deeply bound $ud\bar{b}\bar{b}$ and
$\ell s\bar{b}\bar{b}$ states.

Key features that distinguish our calculation from previous lattice explorations are
the use of NRQCD, thus avoiding the static approximation for the heavy $b$ quarks,
and the use of ensembles with light quarks close to the physical point.

\section{Lattice operators and correlators}

The general form of a Euclidean time, lattice QCD correlation function is (see e.g. \cite{Gattringer:2010zz}) 
\begin{equation}
\begin{aligned}
C_{\mathcal{O}_1\mathcal{O}_2}(p,t) &= \sum_x e^{ip\cdot x} \langle \mathcal{O}_1(x,t) \mathcal{O}_2(0,0)^\dagger \rangle\;,\\
&= \sum_n \langle 0 | \mathcal{O}_1 | n \rangle \langle n | \mathcal{O}_2 | 0 \rangle e^{-E_n(p)t}\;,
\end{aligned}
\end{equation}
where the operators ($\mathcal{O}_i$) have the quantum numbers of the continuum 
state of interest. For the $I(J^P)={\frac{1}{2}}(0^-)$ and ${\frac{1}{2}}(1^-)$
$B(5279)$ and $B^*(5325)$ mesons, for example \footnote{Greek indices here denote spin, 
Latin indices color and $u,d,s,b$ quark flavors.},
\begin{equation}\label{eq:pp_vv}
\begin{aligned}
P(x) &= \bar{b}^\alpha_a(x) \gamma_5^{\alpha\beta} u^{\beta}_a(x) \;,\:\\
V(x) &= \bar{b}^\alpha_a(x) \gamma_i^{\alpha\beta} d^{\beta}_a(x) \;.
\end{aligned}
\end{equation}

We are focused on the $\bar{3}_F$, $J^P=1^+$ $ud\bar{b}\bar{b}$ and $\ell s\bar{b}\bar{b}$
channels, and aim to construct lattice interpolating operators having
good overlap with the expected tetraquark ground states.  

Our first operator has the favorable diquark-antidiquark structure noted above, with 
$\bar{b}\bar{b}$ color $3_c$, spin $1$ and light quark flavor-spin-color 
$(\bar{3}_F,0,\bar{3}_c)$ \footnote{We have used the identity $\epsilon_{eab}\epsilon^{ecd}=\delta_{ac}\delta_{bd}-\delta_{ad}\delta_{bc}$; the two terms on the RHS yield identical contributions to the final correlator.}:
\begin{equation}\label{eq:di_antidi}
\begin{aligned}
D(x) = ( u^\alpha_a(x) &)^T ( C\gamma_5 )^{\alpha\beta} q^{\beta}_b(x) \; \\
& \times\;\bar{b}^\kappa_a(x) ( C\gamma_i )^{\kappa\rho} ( \bar{b}^\rho_b(x) )^T\;,
\end{aligned}
\end{equation}
where $q =d$ or $s$. Though there can be relative orbital momentum, the ground state should have none, yielding a $J^P=1^+$ state. In general, $D(x)$ will also couple
to any pair of conventional mesons with the same quantum numbers (the lowest lying being
$BB^*$ with $L=0$ for $q =d$ and $B_sB^*$ with $L=0$ for $q =s$). Combining a pair of heavy-light mesons on the lattice we are led to consider a meson-meson operator
\begin{equation}\label{eq:dimeson}
\begin{aligned}
M(x) = 
&\, \bar{b}^\alpha_a(x) \gamma_5^{\alpha\beta} u^{\beta}_a(x) \  
\bar{b}_b^{\kappa}(x) \gamma_i^{\kappa\rho} d_b^{\rho}(x)\\
&\, - \bar{b}^\alpha_a(x) \gamma_5^{\alpha\beta} d^{\beta}_a(x)\  
\bar{b}_b^{\kappa}(x) \gamma_i^{\kappa\rho} u_b^{\rho}(x)\;
\end{aligned}
\end{equation}
for the $\bar{3}_F$, $I=0$ channel, and the analogous operator with $B_sB^*$ structure for the $\bar{3}_F$ isodoublet channel.

To study possible tetraquark binding, we 
compare the ground state and lowest-lying free two-heavy-light meson state mass
sum in the channel of interest. This can be achieved by using 
the relevant pseudoscalar (P) and vector (V) meson correlators,
$C_{PP}(t)$ and $C_{VV}(t)$ of Eq.~\ref{eq:pp_vv}, to 
compute the binding correlator,
\begin{equation}\label{eq:binding}
G_{\mathcal{O}_1 \mathcal{O}_2}(t) = \frac{C_{\mathcal{O}_1 \mathcal{O}_2}(t)}{C_{PP}(t) C_{VV}(t)}\;,
\end{equation}
which, for a channel with a tetraquark ground state with (negative) binding
$\Delta E$ with respect to its two-meson PV threshold, grows as
$e^{-\Delta E\, t}$ for large Euclidean $t$.

\subsection{GEVP analysis}

The operators (Eqs.~\ref{eq:di_antidi} and \ref{eq:dimeson}) have the same quantum numbers and hence overlap with the same ground and excited states, though with
different relative strengths. We define the matrix of binding correlation functions,
including possible operator mixing, by
\begin{equation}
F(t) = 
\begin{pmatrix}
G_{DD}(t) & G_{DM}(t)  \\
G_{MD}(t) & G_{MM}(t) 
\end{pmatrix}\;.
\label{eq:gmatrix}
\end{equation}
The variational method can then be used to extract the binding by solving the GEVP,
\begin{equation}
F(t)\nu = \lambda(t) F(t_0) \nu\;,
\label{eq:gevp}
\end{equation}
with the eigenvectors $\nu$ and the binding energy determined directly from the eigenvalues $\lambda(t)$ via,
\begin{equation}
\lambda(t) = A\,e^{-\Delta E ( t - t_0 ) } = ( 1 + \delta )e^{-\Delta E ( t - t_0 ) }\;.
\label{eq:eigval}
\end{equation}
From a $2\times2$ matrix two eigenvalues can be extracted; one corresponds to the ground state and the other to a mixture of all excited state contaminations. 

\begin{table}[t!]
\centering
\begin{tabular}{c|ccc}
\hline\hline
Label & $E_H$ & $E_M$& $E_L$ \\
Extent & $\:32^3\times64\:$ & $\:32^3\times64\:$ & $\:32^3\times64\:$ \\
$a^{-1}\;\left[\text{GeV}\right]$\cite{Namekawa:2013vu} & 2.194(10) & 2.194(10) & 2.194(10) \\
$\kappa_l$ & $0.13754$ & $0.13770$ & $0.13781$ \\
$\kappa_s$ & $0.13640$ & $0.13640$ & $0.13640$ \\
$am_\pi$ & 0.18928(36) & 0.13618(46) & 0.07459(54) \\
$am_K$ & 0.27198(28) & 0.25157(30) & 0.23288(25) \\
$m_\pi L$ & 6.1 & 4.4 & 2.4 \\
$M_{\Upsilon}\;\left[\text{GeV}\right]$ & 9.528(79) & 9.488(71) & 9.443(76)\\
Configurations & 400 & 800 & 195 \\
Measurements & 800 & 800 & 3078 \\
\hline\hline
\end{tabular}
\caption{{Overview of our ensemble parameters: $am_{\pi,K}$ are from global 
cosh/sinh fits to a shared mass and common amplitudes over the $C_{PP}$, $C_{A_tA_t}$, and $C_{A_tP}$ correlators using both wall-local and wall-wall data. Fit ranges were chosen so
$\chi^2/dof$ is close to 1. This analysis leads to $am_\pi$ with uncertainties improved
by a factor of $\sim 6$ relative to those of~\cite{Aoki:2008sm}. Throughout the strange quark is tuned to its physical value in the valence sector with $\kappa_s^{\rm val}=0.13666$. These configurations 
use the Iwasaki gauge action~\cite{Iwasaki:1985we} with $\beta=1.9$ and non-perturbative clover 
coefficient $c_{SW}=1.715$. 
}}
\label{tab:lat_par}
\end{table}

\section{Numerical setup}

We use $n_f=2+1$ Wilson-Clover \cite{Sheikholeslami:1985ij} fermion gauge field ensembles generated by the PACS-CS collaboration \cite{Aoki:2008sm}, with a partially-quenched valence strange quark tuned to obtain the physical $K$ mass at the physical point.
An overview of the ensembles can be found in Tab.~\ref{tab:lat_par}. The basic spectrum of \cite{Aoki:2008sm} was reproduced in this work.
In the valence sector we used Coulomb gauge-fixed wall sources \footnote{The FACG algorithm of \cite{Hudspith:2014oja} was used to fix to an accuracy of $\Theta<10^{-14}$.}. 
Sources on multiple time-source positions were inverted to compute light, strange and bottom quark propagators \footnote{We use a modified deflated SAP-solver \cite{Luscher:2005rx} for the light and strange quarks.}.

\subsection{NRQCD propagators and mass tuning}

We use the NRQCD lattice action \cite{Thacker:1990bm, *Lepage:1992tx, *Manohar:1997qy} to calculate bottom quark propagators. The Hamiltonian is \cite{Davies:1994mp, *Lewis:1998ka, *Lewis:2008fu}
\begin{equation}
\begin{aligned}
H = &-\frac{\Delta^{(2)}}{2M_0}-c_1\frac{(\Delta^{(2)})^2}{8M_0^3}
+ \frac{c_2}{U_0^4}\frac{ig}{8M_0^2}(\bm{\tilde\Delta\cdot\tilde{E}}-\bm{\tilde{E}\cdot\tilde\Delta}) \\
& - \frac{c_3}{U_0^4}\frac{g}{8M_0^2}\bm{\sigma\cdot}(\bm{\tilde\Delta\times\tilde{E}}-\bm{\tilde{E}\times\tilde\Delta})\\
&-\frac{c_4}{U_0^4}\frac{g}{2M_0}\bm{\sigma\cdot\tilde{B}}
+ c_5\frac{a^2\Delta^{(4)}}{24M_0}
- c_6\frac{a(\Delta^{(2)})^2}{16nM_0^2}\;,
\end{aligned}
\label{eq:nrqcdhamilton}
\end{equation}
with the tadpole-improvement coefficient $U_0$ set via the fourth root of the plaquette and tree-level values $c_i=1$. A tilde denotes tree-level improvement and the $c_5,c_6$ terms remove the remaining $\mathcal{O}(a)$ and $\mathcal{O}(a^2)$ errors.

To tune the bottom quark bare mass we calculated the $\Upsilon$-meson correlation function using local-local hadron correlators at finite momentum. The tuning is implemented via the momentum dispersion relation
\begin{equation}\label{eq:dispersion}
E(\hat{p})=M_0 + \frac{1}{2}\frac{\hat{p}^2}{M_{\rm ph}} + \sum_{n>1}\mathcal{O}(\hat{p}^{[2n]}),
\end{equation}
where we use lattice momenta $\hat{p}_\mu=a\sin\left(\frac{2\pi n_\mu}{L_\mu}\right)$. $M_{\rm ph}$ is the physical hadron mass and the values quoted in Tab.~\ref{tab:lat_par} are from a linear fit in $\hat{p}^2$ to Eq.~\ref{eq:dispersion}.
This setup is known to account for relativistic effects at the few percent level while capturing the relevant heavy-light quark physics \cite{Gray:2005ur,Brown:2014ena,*Lewis:2008fu} \footnote{Our own heavy meson and baryon spectrum agrees well with \cite{Brown:2014ena} and a publication is in preparation.}. 

\section{Numerical results, chiral and volume extrapolations}

\begin{figure}
 \centering
 \includegraphics[width=0.48\textwidth]{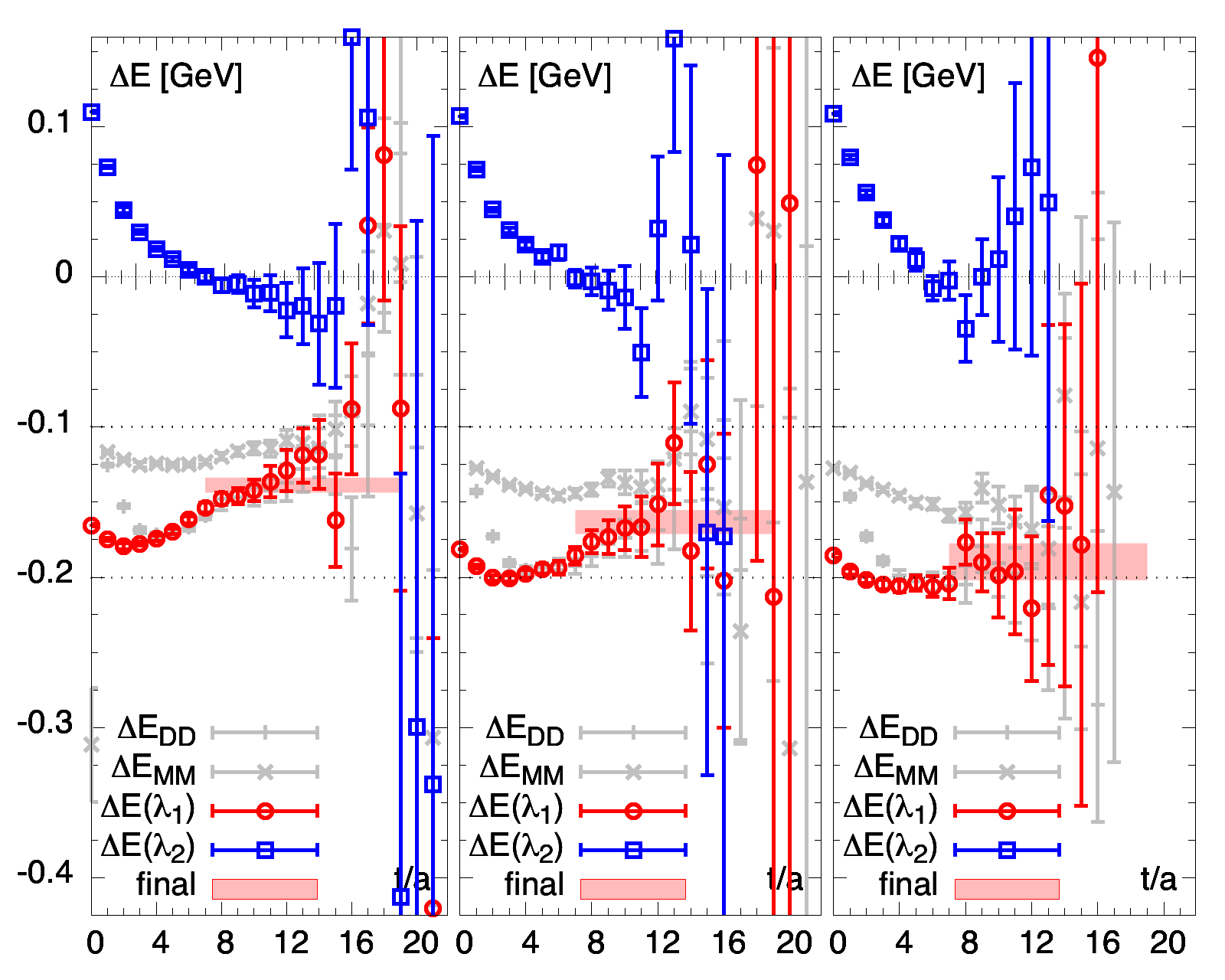}
 \includegraphics[width=0.48\textwidth]{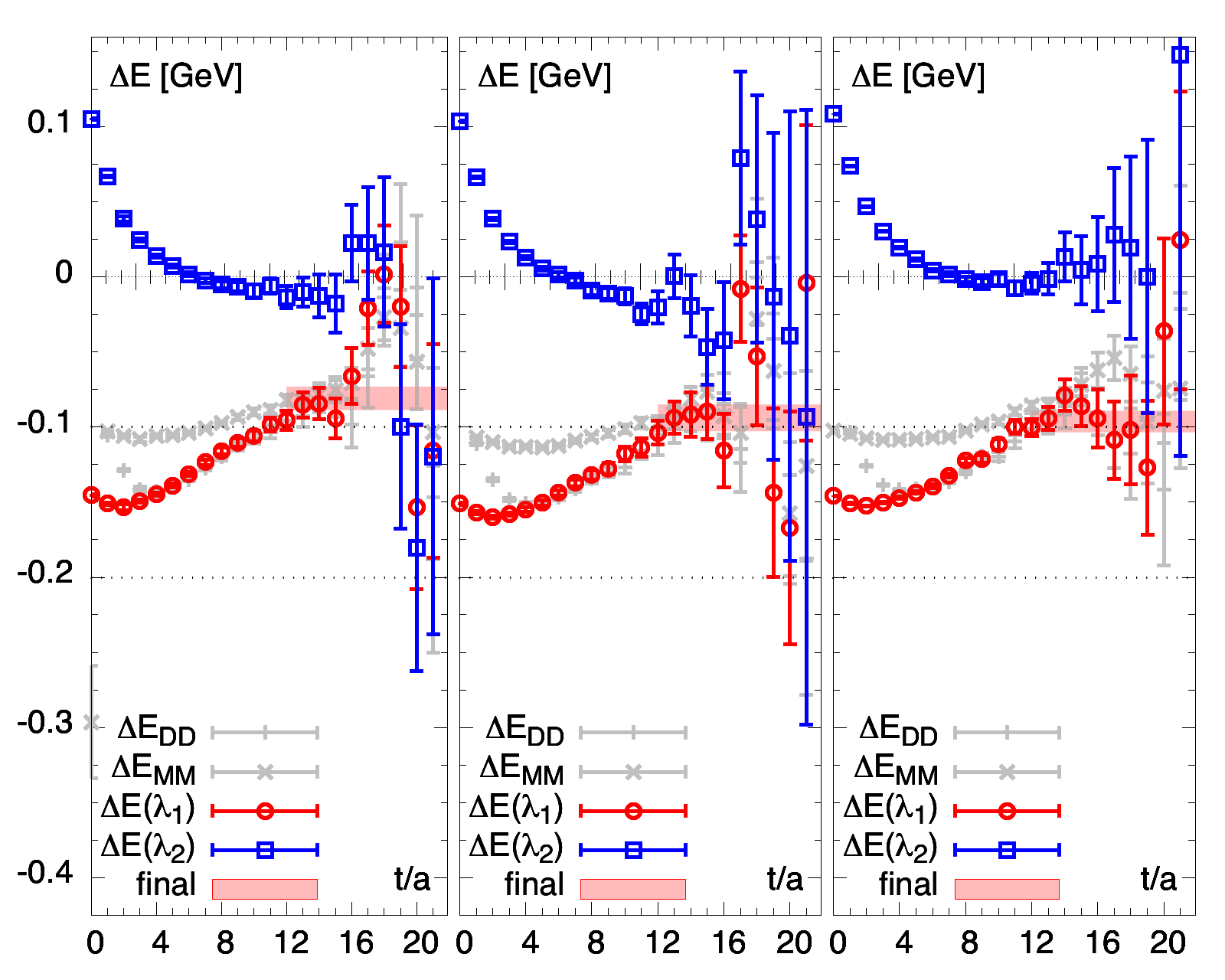}
 \caption{$ud\bar{b}\bar{b}$ (top panel) and $ \ell s\bar{b}\bar{b}$ (bottom panel) tetraquark effective binding energies. Red circles (blue squares) represent the bindings
relative to the $BB^*$ ($B_sB^*$) threshold of the first and second GEVP eigenvalues, respectively. Red bands denote the final fit results. Grey dashes and grey crosses indicate the 
bindings obtained from the corresponding diquark-diquark and meson-meson single-operator analyses. Left panels: $E_H\,(m_\pi L\simeq 6.1,~m_\pi\simeq 415~$MeV). Center: $E_M\,(m_\pi L\simeq 4.4,~m_\pi\simeq 299~$MeV). Right: $E_L\,(m_\pi L\simeq 2.4,~m_\pi\simeq 164~$MeV).}
 \label{fig:llbb_eig}
 \end{figure}

Our results for the ground (red) and excited state (blue) binding energies are shown in Fig.~\ref{fig:llbb_eig}. For presentational purposes we show these as log-effective binding energies. For comparison,
results obtained from the single-operator diquark-antidiquark (grey dashes) and meson-meson (grey crosses) analyses are also included. The results show that both operators couple well to the ground state. We also see, as $t/a$ increases, the second GEVP eigenvalue 
approach the relevant two-meson PV threshold in both channels, strongly supporting 
an interpretation of the corresponding ground states as genuine tetraquarks \footnote{In addition, we estimated the effect of a possible attractive meson-meson interaction for a hypothetical $BB$-system using the finite volume relations of \cite{Luscher:1986pf} and find it to be at the $\Delta E\approx -10$MeV level for both the ground and threshold energies. }.

To estimate the binding energy we perform a single exponential fit, Eq.~\ref{eq:eigval}, to the first eigenvalue $\lambda(t)$ and accept those that satisfy $\chi^2/d.o.f. \sim 1$. In case of an increasing exponential in time, which would indicate a state below threshold, the quality of $\chi^2/d.o.f.$ diminishes as more noise dominated points are added at long distances. We observe this effect and, in order to give conservative estimates, for our final results we choose the longest fit range with $\chi^2/d.o.f.\simeq 1$ in $t/a$; these are $7\rightarrow 19$ and $12\rightarrow 25$ for the $ud\bar{b}\bar{b}$ and $\ell s\bar{b}\bar{b}$ channels, respectively.

\begin{figure}
\centering
\includegraphics[width=0.48\textwidth]{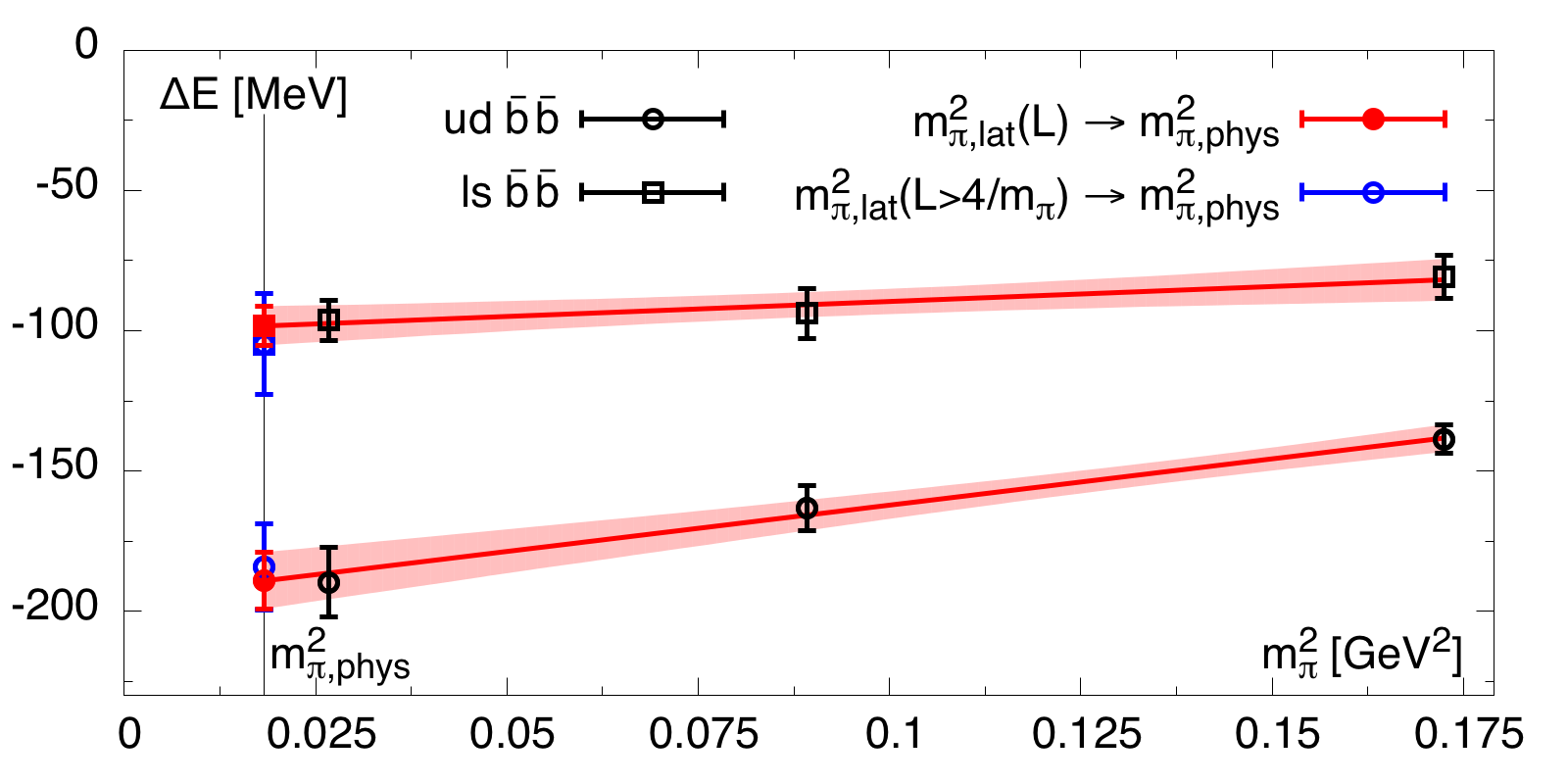}
\caption{Chiral extrapolations of the $ud\bar{b}\bar{b}$ and $\ell s\bar{b}\bar{b}$ binding energies. Red lines and points show the extrapolations using all 
three ensembles, the blue points those using $E_H$ and $E_M$.}
\label{fig:chiral}
\end{figure}
We use a linear extrapolation in  $m_\pi^2$ to determine our physical point tetraquark bindings \footnote{This is the leading order chiral behavior when the strange quark masses on all ensembles have been tuned to the physical value \cite{Cheng:1993kp}, as was done here.}.
As the ensemble $E_L$ has a small $m_\pi L$, we estimate our finite volume and chiral extrapolation systematic by performing two such extrapolations, one using only 
$E_H$ and $E_M$ and the other using all three ensembles, taking half the difference of the resulting central values as our systematic error. These extrapolations are shown in Fig.~\ref{fig:chiral}, with the filled red symbols giving the 
physical point results for the three-ensemble fits and the open blue 
symbols the corresponding results for the fits employing only $E_H$ 
and $E_M$. The results of both extrapolations are in
good agreement, implying that finite volume errors are under control. 
The individual-ensemble and extrapolated physical-point 
results are given in Tab.~\ref{tab:results}.
Light quark cut-off effects are at the $\mathcal{O}(a^2)$-level and hence expected to be small, while the NRQCD Hamiltonian in Eq.~\ref{eq:nrqcdhamilton} is $\mathcal{O}(a^2)$ improved.

\begin{table}
\centering
\begin{tabular}{ccc}
\toprule
Ensemble & $\Delta E_{ud\bar b \bar b}$[MeV] & $\Delta E_{\ell s\bar b \bar b}$[MeV] \\
\noalign{\smallskip}\hline\noalign{\smallskip}
$E_H$ & -139(5) & -81(8) \\
$E_M$ & -163(8) & -94(9) \\
$E_L$ & -190(12) & -96(7) \\
Phys & -189(10)(3) & -98(7)(3) \\
\hline
\bottomrule
\end{tabular}
\caption{Ensemble and extrapolated physical-point (Phys) $ud\bar{b}\bar{b}$
and $\ell s\bar{b}\bar{b}$ binding energies from fitting all ensembles. Errors for the 
individual ensembles are statistical. For the extrapolated physical
point entries, the first error is statistical and the second 
the systematic error estimated as described in the text.}
\label{tab:results}      
\end{table}

\section{Decay modes suitable for experimental detection}

We discuss briefly decay modes likely to be amenable to experimental searches 
for the $\bar{3}_F$, $J^P=1^+$ $qq^\prime \bar{b}\bar{b}$ tetraquark candidates identified above. 

With a binding of $189$ MeV relative to its $BB^*$ strong-interaction-stability 
threshold, a $ud\bar{b}\bar{b}$ tetraquark will lie below $BB$ threshold, 
and hence also be stable with respect to electromagnetic 
decays. The same is true of a $\ell s\bar{b}\bar{b}$ tetraquark bound by $98$ MeV. 
With both $ud\bar{b}\bar{b}$ and $\ell s\bar{b}\bar{b}$ tetraquarks decaying only weakly, the resulting displaced decay vertices 
should aid in searching for these states experimentally.

Examples of fully reconstructable modes for the weak decay of the $ud\bar{b}\bar{b}$ 
tetraquark are $B^+\bar{D}^0$ and $J/\Psi B^+ K^0$, with
$\bar{D}^0$ and $B^+$ fully reconstructable from $\bar{D}^0\to K^+\pi^-$,  
$B^+\to\bar D^0\pi^+$, and $K^0$ from its $\pi^+\pi^-K_S$ decay.
Similarly, $J/\Psi B_s K^+$ and $J/\Psi B^+\phi$ would serve as 
fully reconstructable modes for the weak decay of the
$us\bar{b}\bar{b}$ tetraquark, and $B^+ D_s^-$, $B_s\bar{D}^0$, 
$J/\Psi B^0\phi$ and $J/\Psi B_s K^0$ for the $ds\bar{b}\bar{b}$ tetraquark.

\section{Conclusions}

We predict the existence of a $\bar{3}_F$ of strong- and electromagnetic-interaction
stable $qq^\prime\bar{b}\bar{b}$ tetraquarks with $ud\bar{b}\bar{b}$ and 
$\ell s\bar{b}\bar{b}$ member masses $10.415(10)$ and $10.594(8)$ GeV, respectively. 
These states should decay only weakly, 
with ordinary heavy meson decay products emitted from a displaced vertex.

While the doubly bottom nature of these states may make experimental detection challenging, decay modes with favorable experimental tag possibilities do exist, making searches for these states interesting.
Analogous $qq^\prime \bar c\bar c$, $qq^\prime \bar c\bar b$, and $qq^\prime \bar s \bar b $ tetraquarks, if also stable with respect to both strong
and electromagnetic decays, would be more easily detectable experimentally. Whether or not such lighter tetraquark states exist is not clear at present, but is the subject of ongoing investigations, the results of which will be reported in a future publication.


\section*{Acknowledgments}
We thank Mark Wurtz for help with the NRQCD portion of this work.
The authors are supported by NSERC of Canada. Propagator inversions and gauge fixing were performed on Compute Canada's GPC machine at SciNet. Contractions were performed using our open-source contraction library \cite{contractions}.

\bibliography{tetra.bib}

\end{document}